\documentstyle[aps,preprint,multirow,hhline,epsfig,amsfonts]{revtex}     

\begin{document}   
   
\title{Information processing by a controlled coupling process}

\author{G. W. Wei and Shan Zhao}

\address{Department of Computational Science,\\    
        National University of Singapore,   
        Singapore 117543, Singapore}    

\date{\today}     
\maketitle   
\begin{abstract}
This Letter proposes a controlled coupling process for 
 information processing.
The net effect of conventional coupling is isolated 
from the dynamical system and is analyzed in depth.
The stability of the process is studied. We show that 
the proposed  process  can locally minimize 
the smoothness and the fidelity of dynamical data.
A digital filter expression of 
the controlled coupling process is derived and
the connection is made to the Hanning filter. 
The utility and robustness of proposed approach is 
demonstrated by both the restoration of the contaminated
solution of the nonlinear Schr\"{o}dinger equation 
and the estimation of the trend of 
a time series.
\end{abstract}
PACS numbers: 05.45.Gg, 05.45.Tp, 05.45.Xt

\vskip 12pt

The topic of synchronization and chaos control has attracted 
much attention in the past 
decade\cite{P1,KocPar,AVPS,FCR,Ritort,SchMar,Weiprl,LMDIB,YHX}.
Active research in this area has contributed greatly 
to the understanding of a wide class of complex phenomena, 
including synchronization in secure communication\cite{KocPar},
electronic circuits\cite{AVPS}, and nonlinear optics\cite{FCR},
coherence transfer in magnetic resonance\cite{Ritort}, and
oscillation in chemical and biological systems\cite{SchMar}.
In fact, the study of this topic leads to many practical 
applications in the aforementioned fields and a realistic 
scheme for shock capturing\cite{Weiprl} in fluid dynamics. 
An interesting signal processing scheme
was proposed by Lindner {\it et. al.}\cite{LMDIB}
in the context of synchronization.
It is beneficial to explore further the potential of 
synchronization and chaos control techniques to 
digital signal processing (DSP) and information autoregression (IAR). 
Both  DSP and IAR are of crucial importance 
to telecommunication, biomedical imaging, seismic,  
radar, pattern recognition,  missile guidance, target tracking, 
autonomous control, and a wide variety of other signals and data
processing for commerce, finance, defense and scientific interests. 
Despite of the great achievements of traditional DSP  
and IAR techniques, many real-world problems remain  
unsolved due to the nonlinearities, non-Gaussian noises and  
limited real-time working conditions often  
encountered in  applications. 
Solving these  complex problems would require the development of  
innovative, flexible, fast, nonlinear DSP and IAR algorithms.
The objectives of this Letter is to develop a 
synchronization-based, realistic technique for DSP and IAR.

\begin{figure*}[!htb]  
\begin{center}  
    \leavevmode     
      \includegraphics[width=0.70\textwidth]{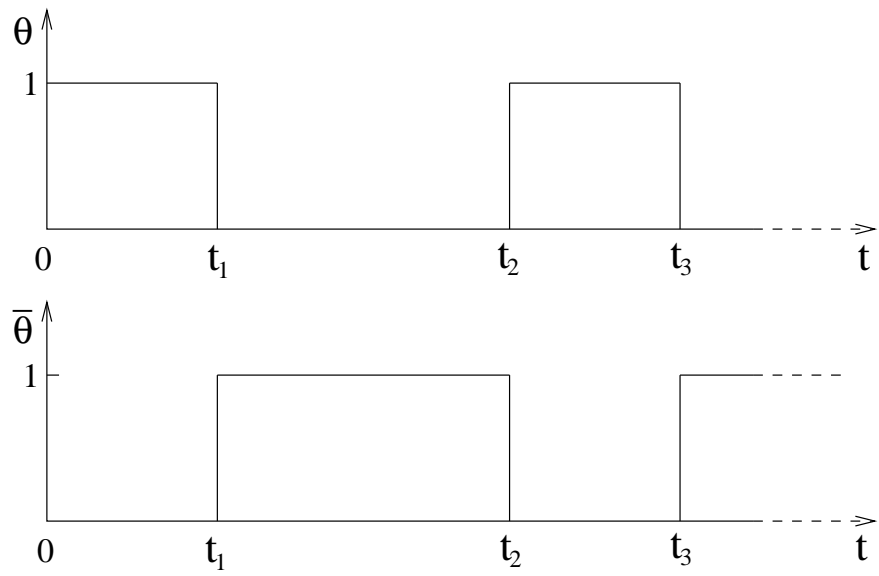}  
\caption{The control functions.}
\label{fig.coupling}  
\end{center}  
\end{figure*}

We consider a nonlinear dynamical system consisting of 
$N$ identical subsystems, which are coupled via   
the nearest and next-to-the-nearest neighborhood sites
\begin{eqnarray}\label{synch1}  
{du_{j}\over dt} &=& f(u_{j}) +N_j(t)\nonumber \\
&+&a(u_{j+2}-u_{j+1})
+(b-3a)(u_{j+1}-u_{j}) \nonumber \\
&+& a (u_{j-2}-u_{j-1})
+(b-3a)(u_{j-1}-u_{j}),
\end{eqnarray}  
where $u_j\in [0,\infty)\times R^n$, 
$f$ is a nonlinear function of $u_j$ which 
might undergo chaotic dynamics, $N_j(t)$ is the noise,
$a$ and $b$ are scalar hyper-diffusive
and diffusive coupling parameters, respectively.
Equation  (\ref{synch1}) is a generalization of that 
given by  Lindner {\it et. al.}\cite{LMDIB}.
The coupling scheme in Eq. (\ref{synch1}) is 
strongly dissipative and a synchronous state can be 
attained for the  chaotic Duffing oscillators, 
$\dot{u}=(\dot{x}, \dot{y})= (y,-0.3y-x^3+11\cos t)$,
by using appropriate coupling parameters\cite{YHX}.
It is important to analyze in depth the effect
of the coupling in Eq. (\ref{synch1}).
To this end, we introduce the following 
controlled coupling process
\begin{eqnarray}\label{synch2}  
{du_{j}\over dt} &=& 
\theta_{t_1,t_2,\cdots,t_k}(t)\left[ f(u_{j})+ N_j(t)\right] \nonumber \\
&+&\bar{\theta}_{t_1,t_2,\cdots,t_k}(t)
\left[ a(u_{j+2}-u_{j+1})
+(b-3a)(u_{j+1}-u_{j})\right. 
\nonumber \\
&+& \left.
a(u_{j-2}-u_{j-1})
+(b-3a)(u_{j-1}-u_{j})\right],
\end{eqnarray}  
where $\theta_{t_1,t_2,\cdots,t_k}(t)$ is a control function
which consists of a train of Heaviside type intervals and 
$\bar{\theta}_{t_1,t_2,\cdots,t_k}(t)$ is the complement of 
$\theta_{t_1,t_2,\cdots,t_k}(t)$ in the domain $[0,\infty)$ 
[i.e., $\bar{\theta}_{t_1,t_2,\cdots,t_k}(t)=1-\theta_{t_1,t_2,
\cdots,t_k}(t)$]. Both control functions are depicted in FIG. 
\ref{fig.coupling}.
During the time interval $0 \leq t \leq t_1$, each subsystem,
including possible noise,
evolves freely. At time $t_1$, the coupling is switched on 
and the time evolution of the system is entirely governed by 
the controlled coupling until $t_2$. The subsystems 
return to the state of free evolution in the next time interval.

The controlled coupling process, Eq. (\ref{synch2}), allows
a detailed analysis of the {\it net effect} of the coupling 
in Eq. (\ref{synch1}). 
At time $t_1 \leq t \leq t_2$, 
a discrete form of Eq. (\ref{synch2}) can be given by 
\begin{eqnarray}\label{jump.pro}  
u^{S+1}_{j} & = & u^{S}_{j}  
+ R(u^{S}_{j-1}-2u^{S}_{j}+u^{S}_{j+1})\nonumber\\   
&+& T(u^{S}_{j-2}-4u^{S}_{j-1}
+6u^{S}_{j}-4u^{S}_{j+1}+u^{S}_{j+2}),\\ 
u^{0}_{j} & = & u_{j}(t_1)  \qquad j=1,\ldots,N, \nonumber  
\end{eqnarray}  
where $R=b\Delta t $, $T=a\Delta t $
and iteration parameter $S$ are user-specified  
constants. 
The controlled coupling  process (\ref{jump.pro}) is
conditionally stable. 
Neglecting the boundary modifications, we can rewrite Eq.  
(\ref{jump.pro}) in a matrix form,  
\begin{equation}\label{JP.matrix}  
U^{S+1} = A U^{S},  
\end{equation}  
where $U^{S}=(u^{S}_{1},u^{S}_{2},\ldots,u^{S}_{N})^{\rm T}$, 
and the banded matrix $A$ has nonzero coefficients: 
$a_{j,j-2}=a_{j,j+2}=T$, 
$a_{j,j-1}=a_{j,j+1}=R-4T$,
and $a_{j,j}=1-2R+6T$, for $j=1,2,\ldots,N$. 
If all of the eigenvalues of $A$ are  
smaller than unity, the iterative correction $\epsilon^{S+1}=|| U^{S+1}-   
U^{S} ||$ will decay, then the process is stable. 
Since each diagonal term of the matrix is a constant,
the eigenvectors of $A$ can  
be represented  in terms of a complex exponential form, 
\begin{equation}\label{eigen.vt}  
U^{S}_{j} = q^{S} e^{i \gamma j}, 
\end{equation}  
where $i=\sqrt{-1}$ and $\gamma$ is a wavenumber that can be chosen  
arbitrarily. Substituting Eq. (\ref{eigen.vt}) into Eq. (\ref{JP.matrix}) and  
removing the common term $e^{i \gamma j}$, we obtain an explicit expression  
for the eigenvalue $q$:  
\begin{equation}  
q = 1 + 2T(2\cos^2 \gamma -\cos \gamma -1)+
2(R-3T)(\cos \gamma -1).  
\end{equation}  
For a stable process, the magnitude of this quantity is required to be  
 smaller than unity, 
\begin{equation}  
q^{2} < 1. 
\end{equation} 
For the case of $T=0$,
$q$ is the maximum when $\cos \gamma = -1$. 
Thus, the controlled coupling
process is stable provided $0<R<\frac{1}{2}$.   
On the other hand, if $R=0$,
our analysis indicates that 
$0> T >- {1\over 2}{1\over (\cos^2 \gamma -2\cos \gamma+1)}$, which 
also takes  an extremum at $\cos \gamma=-1$. Therefore, 
the controlled coupling
process is stable provided $0>T>-\frac{1}{8}$.
Under these conditions, we have 
$\epsilon^{S+1} \le \epsilon^{S}$, for any $S \in \mathbb{Z}^{+}$.

It is interesting to noted that  the controlled coupling 
process, Eq. (\ref{synch2}), provides not only a treatment of 
nonlinear dynamical systems, but also a powerful, realistic 
approach to DSP and IAR of real-world problems. 
For IAR, Whittaker \cite{Whitta} suggested a method of 
trend estimation by the global minimization of fidelity and 
smoothness. The latter is defined by the accumulated 
power of the finite difference. Hodrick and Prescott\cite{HodPre} 
provided a concrete version of Whittaker's approach.
Recently, Mosheiov and Raveh \cite{MosRav} proposed a linear   
programming approach to estimate the trend by employing the sum of the  
{\em absolute} values rather than the common sum of squares to 
measure  the smoothness and fidelity. In the present 
approach, the terms in the first and second brackets  
of Eq. (\ref{jump.pro}) are the second order and fourth 
order pointwise measures of smoothness, which are denoted 
as $\Delta^{2} u^{S}_{j}$ and $\Delta^{4} u^{S}_{j}$, respectively.  
To have a better understanding of this controlled coupling
process, we rewrite Eq. (\ref{jump.pro}) as 
\begin{equation}\label{JP.cret}  
[ u_{j}(t_1) - u^{S}_{j} ]+ [R v^{S-1}_{j} +T w^{S-1}_{j}] 
=0, \quad j=1,\ldots,N,  
\end{equation}  
where $v^{S-1}_{j} = \sum_{k=0}^{S-1} \Delta^{2} u^{k}_{j}$,   
and $w^{S-1}_{j} = \sum_{k=0}^{S-1} \Delta^{4} u^{k}_{j}$.   
It is clear that the expression in the first square 
bracket is the local measure of the fidelity, 
while the expression in the second square bracket
is the accumulative local measures of smoothness. 
Due to $\epsilon^{S+1}=|| U^{S+1}- U^{S} || = ||   
R\Delta^{2} U^{S} + T\Delta^{4} U^{S} ||$, 
is actually a global smoothness measure of  
estimated trend at the $S$th iteration. One can argue that as 
the iterative process is carried out for a longer time,
the estimated trend becomes smoother, while the deviation of  
$U^{S}$ from $U^0=U(t_1) 
=[u_{1}(t_1),u_{2}(t_1),\ldots,  
u_{N}(t_1)]^{\rm T}$ 
becomes larger.
At  each step of the  iteration, this process guarantees 
that the sum of the local deviation from $u_{j}(t_1)$ and the  
accumulative local measure of smoothness equals to zero.
As such, the result of each iteration is optimal 
in the sense of minimization, for the given input and 
the set of parameters $R$ and $T$. Two smoothing parameters 
$R$ and $T$, and the iteration parameter $S$, govern the 
fundamental tradeoff between the smoothness and   
fidelity. In practice, $R$ and/or $T$ can be pre-fixed 
and only the iteration parameter $S$ is optimized to achieve 
desired results.  In comparison, the previous IAR methods seek 
for global minimizations over the entire  domain to obtain 
optimal estimates, while the present controlled 
coupling process forces the sum of smoothness and   
fidelity to pass through zero at each iteration 
to give an optimal trend. The advantage of the proposed 
controlled coupling is its simplicity, robustness and 
efficiency.

\begin{figure*}[!htb]  
\begin{center}  
    \leavevmode     
\includegraphics[width=0.80\textwidth]{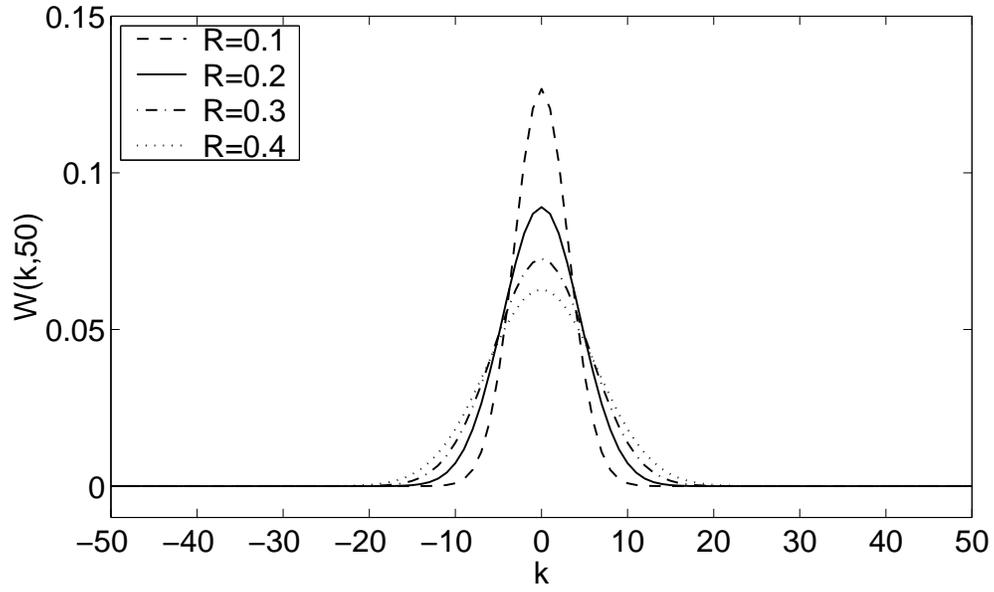}
\caption{The filter weights of the controlled coupling  
process  ($T=0$).}  
\label{fig.weight}  
\end{center}  
\end{figure*}

For DSP, it is important to analyze
the relationship between the proposed process
and digital filters. To this end, 
we explore a weighted average representation 
of the controlled coupling  process.
For simplicity, we consider the case of $T=0$.
We first set $S$ to $1$, then the controlled 
coupling process gives
\begin{equation}\label{JP.wa_1}  
u_{j}^{1} = R u_{j-1}(t_1)  + 
(1-2R) u_{j}(t_1) + R u_{j+1}(t_1), \quad j=1,\ldots,N,  
\end{equation}  
which is clearly a local weighted average form for $u_{j}(t_1)$. 
In general, after $S$ iterations, the controlled coupling 
process can be  represented as:  
\begin{equation}\label{JP.wa_m}  
u_{j}^{S} = \sum_{k=j-S}^{j+S} W(k,S) u_{k}(t_1) , 
\end{equation}  
where weight function $W(k,S)$ has the general form of 
\begin{equation}\label{JP.weight}  
W(k,S) = \left\{ \begin{array}{lll}  
\sum_{h=0}^{(S-k)/2}  & g(k,S,2h) & \textrm{when}~S-k~\textrm{even}\\  
\sum_{h=1}^{(S-k+1)/2} &  g(k,S,2h-1) & \textrm{when}~S-k~\textrm{odd}, 
\end{array} \right. 
\end{equation}  
and  
\begin{equation}  
g(k,S,h) = \frac{S! R^{S-h} (1-2R)^{h}}{(\frac{S+k-h}{2})!   
(\frac{S-k-h}{2})! h!}. 
\end{equation}  
It can be easily verified that, 
\begin{equation}  
\sum_{k=j-S}^{j+S} W(k,S)=1,
\end{equation}  
and  
\begin{equation}  
W(-k,S)=W(k,S) \quad \forall k=1,\ldots,S. 
\end{equation}  
Equation (\ref{JP.wa_m}) indicates that the controlled coupling  process  
can be viewed as a kernel smoother for IAR
and  a low-pass filter for DSP.
The implementation of the controlled coupling  process becomes extremely  
simple due to the existence of Eq. (\ref{JP.wa_m}). Therefore, the 
weighted average form (\ref{JP.wa_m}) is numerically very useful.
The weights assignment of the controlled coupling  
process filter is analogous to  
that of other kernel regression methods. 
For a reasonable choice of $R$ and $S$, the greater or smaller
weight will be assigned to the points close or far away from
$u_{j}(t_1)$, respectively, see Table \ref{table.weight} and
FIG. \ref{fig.weight}.
Obviously, the distribution of the weights has a Gaussian 
shape when $S$ is sufficiently large.

A simple moving average filter can be constructed by convolving
the mask $\left(\frac{1}{2},\frac{1}{2}\right)$ with itself $2S$ times.
when $S=1$, such a filter is the Hanning filter\cite{Goodall} 
$\left(\frac{1}{4}, \frac{1}{2}, \frac{1}{4}\right)$.  
In our case, if we set $R=\frac{1}{4}$ in the 
Eq. (\ref{JP.wa_1}), the present 
controlled coupling  process has identical 
filter coefficients as those of the Hanning filter.
Thus, the proposed controlled coupling
process filter can be viewed as a   
generalization of the Hanning filter.   
A similar analysis including a non-vanishing $T$ can be 
carried out.

\begin{figure*}[!htb]  
\begin{center}  
    \leavevmode     
\includegraphics[width=0.80\textwidth]{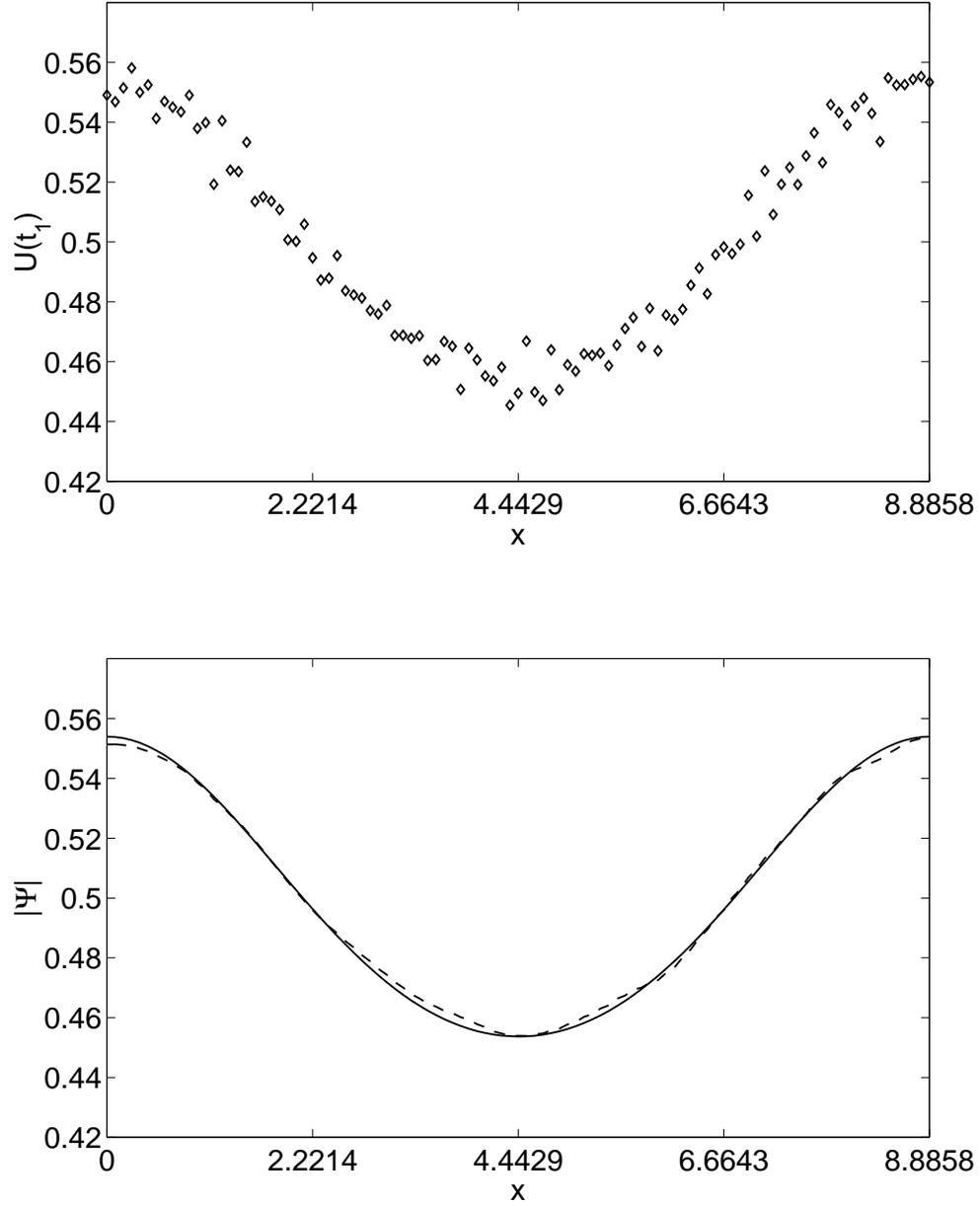}
\caption{Signal restoration from the contaminated solution of 
the nonlinear Schr\"{o}dinger equation.
Upper: the contaminated solution $U(t_1)$; 
Lower: the restored solution (the dashed line)
and  the noise-free solution (the solid line).} 
\label{fig.nls}  
\end{center}  
\end{figure*}

In the rest of this Letter, we demonstrate the utility 
of the proposed approach through numerical experiments.
First, we consider the signal extraction from noisy data by
the controlled coupling process (\ref{synch2}). 
The underlying nonlinear dynamic system is chosen as 
the nonlinear Schr\"odinger equation\cite{HerAbl}
\begin{equation}\label{nlse}
i \frac{\partial \Psi}{\partial t} + \frac{\partial^2 \Psi}
{\partial x^2 } + 2|\Psi|^2 \Psi =0, \quad x\in [0,L],
\end{equation}
with periodic boundary conditions $\Psi(x+L,t)=\Psi(x,t)$
and the period $L=2\sqrt{2}\pi$.  
This system is computationally difficult due to possible
numerically induced chaos\cite{HerAbl}.
In this study, Eq. (\ref{nlse}) is allowed to 
evolves freely with the initial condition of the form 
$\Psi(x,0)=0.5+0.05 \cos\left({2\pi\over L} x\right)+
i10^{-5} \sin\left({2\pi\over L} x\right)$. 
At $t=t_1=8.0$, the system is perturbed and its solution is 
contaminated  by the Gaussian white noise to a 
signal-to-noise ratio (SNR) of $39.25$ dB, 
see FIG. \ref{fig.nls}.
The controlled coupling process is used to restore the solution 
from noisy dynamical data $U(t_1)$  during the time period 
$t_1\leq t \leq t_2$. The parameters
$(R, T, S)$ are chosen as $(0.25, -0.05, 10)$.
From FIG. \ref{fig.nls}, it is 
clear that the unwanted noise is satisfactorily suppressed 
by the proposed process. The restored solution
matches well with the noise-free solution 
and its SNR is as high as $50.63$ dB.

\begin{figure*}[!htb]  
\begin{center}  
    \leavevmode  
\includegraphics[width=0.80\textwidth]{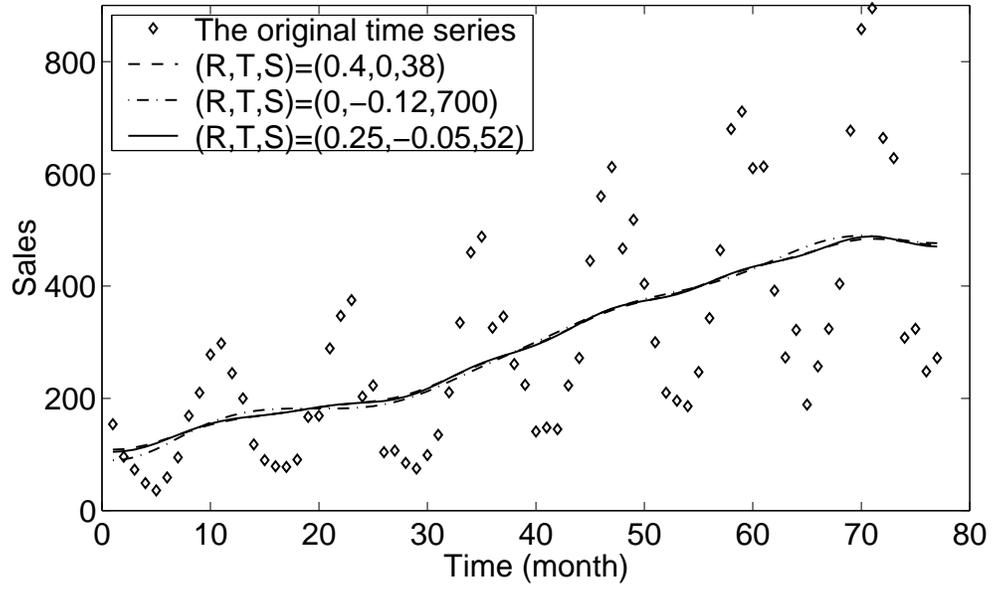}
\caption{Trend estimation by using the controlled coupling process.}
\label{fig.alpha}  
\end{center}  
\end{figure*}

We next consider the trend estimation of a benchmark 
time series, the `Sales of  Company X' series\cite{ChaPro,MosRav}.
Such a time series can be regarded as $U(t_1)$,
produced by an unknown and unpredictable dynamic system,
and its analysis is of practical importance.
The Sales is a monthly series ranging from 
January 1965 to May 1971 and has a monotonic growing trend  
and a clearly identifiable seasonal component. 
Guided by the earlier stability analysis, we choose
three sets of $(R,T)$ values, $(0.4,0)$, $(0,-0.12)$, and 
$(0.25,-0.05)$, for which nearly optimal results
are obtained at $S=38,700$ and 52, respectively, see
FIG. \ref{fig.alpha}. 
The Neumann boundary condition is used to in this case.
As shown in FIG. \ref{fig.alpha},
trends  estimated by using three sets of parameters
are almost identical and provide similar long-run tendency.
An important feature is that,
the slope of the trend undergoes a clear change 
around the $28$th month,  which agrees with the 
finding in Ref. \cite{MosRav}.

In conclusion, we  introduce a controlled coupling process
for the synchronization of spatiotemporal systems.
The proposed process isolates the conventional coupling
from the dynamical system and provides an in-depth 
analysis of the coupling effect. Numerical stability 
of the proposed process is  analyzed.
In the context of trend estimation, 
a comparison is given to several standard
nonparametric methods\cite{Whitta,HodPre,MosRav},
which globally minimization the
smoothness and fidelity.
The proposed process is shown to
balance these features in each step of time evolution,
without resorting to the minimization process,
and thus,  is numerically simpler than the existing methods.
In the context of signal processing,
a digital filter expression of the controlled coupling  
process is derived and the connection of the proposed 
approach to the standard Hanning filter\cite{Goodall} 
is made.  The proposed process is applied to 
signal restoration and trend estimation.

In the numerical experiment of signal restoration, 
the solution of the nonlinear Schr\"odinger equation 
is contaminated by noise at the end of the first time period
 $t_1$. The controlled coupling  process is utilized 
to restore the waveform. We show that the unwanted 
noise can be effectively removed by 10 iterations. 
In the other experiment,  a real-world time series
generated by some unknown dynamical process is studied.
The objective is to estimate the trend of the time series. 
Excellent trend estimations
are obtained by using three different sets of parameters.
The numerical results are in good agreement with 
those in the literature\cite{MosRav} and the present 
approach is simpler. Obviously, the proposed
approach can be easily generalized to two spatial dimensions
for image processing  and the present investigation 
opens up a new opportunity to develop other
synchronization-based, realistic information process 
methods.

This work was supported in part by the National 
University of Singapore and in part by the Center for 
Dynamic Systems and Nonlinear Sciences.

\newpage

\begin{table}[!htb]  
\begin{center}  
\caption{The filter weights [$W(k,6)$] of the controlled 
coupling  process ($T=0$).}  
\label{table.weight}  
\medskip  
\begin{tabular}{c|l|l|l}
~~~$k$~~~ &   & $R=0.4$ & $R=0.1$\\  
\hline  
~~~$0$~~~ & $924R^{6}-1512R^{5}+1050R^{4}-400R^{3}+90R^{2}-12R+1$
& $0.181824$ & $0.390804$\\  
~~~$1$~~~ & $-792R^{6}+1260R^{5}-840R^{4}+300R^{3}-60R^{2}+6R$ 
& $0.154368$ & $0.227808$\\  
~~~$2$~~~ & $495R^{6}-720R^{5}+420R^{4}-120R^{3}+15R^{2}$ 
& $0.12672$  & $0.065295$\\  
~~~$3$~~~ & $-220R^{6}+270R^{5}-120R^{4}+20R^{3}$ 
& $0.07168$  & $0.01048$\\  
~~~$4$~~~ & $66R^{6}-60R^{5}+15R^{4}$ 
& $0.039936$  & $0.000966$\\  
~~~$5$~~~ & $-12R^{6}+6R^{5}$ 
& $0.012288$  & $0.000048$\\ 
~~~$6$~~~ & $R^{6}$ 
& $0.004096$  & $0.000001$\\ 
\end{tabular}  
\end{center}  
\end{table}  

\newpage

\newpage

\newpage

\newpage

\end{document}